\documentstyle[epsfig]{mn}

\newif\ifAMStwofonts
\ifoldfss
  \ifCUPmtlplainloaded \else
    \NewTextAlphabet{textbfit} {cmbxti10} {}
    \NewTextAlphabet{textbfss} {cmssbx10} {}
    \NewMathAlphabet{mathbfit} {cmbxti10} {} % for math mode
    \NewMathAlphabet{mathbfss} {cmssbx10} {} %  "   "    "
  \fi
  \ifAMStwofonts
    \ifCUPmtlplainloaded \else
      \NewSymbolFont{upmath} {eurm10}
      \NewSymbolFont{AMSa} {msam10}
      \NewMathSymbol{\upi}     {0}{upmath}{19}
      \NewMathSymbol{\umu}     {0}{upmath}{16}
      \NewMathSymbol{\upartial}{0}{upmath}{40}
      \NewMathSymbol{\leqslant}{3}{AMSa}{36}
      \NewMathSymbol{\geqslant}{3}{AMSa}{3E}

    \fi
  \fi
\fi % End of OFSS

\ifnfssone
  \newmathalphabet{\mathit}
  \addtoversion{normal}{\mathit}{cmr}{m}{it}
  \addtoversion{bold}{\mathit}{cmr}{bx}{it}
  \newmathalphabet{\mathbfit} % math mode version of \textbfit{..}
  \addtoversion{normal}{\mathbfit}{cmr}{bx}{it}
  \addtoversion{bold}{\mathbfit}{cmr}{bx}{it}
  \newmathalphabet{\mathbfss} % math mode version of \textbfss{..}
  \addtoversion{normal}{\mathbfss}{cmss}{bx}{n}
  \addtoversion{bold}{\mathbfss}{cmss}{bx}{n}
  \ifAMStwofonts
    \ifCUPmtlplainloaded \else
      \UseAMStwoboldmath
      \makeatletter
      \new@mathgroup\upmath@group
      \define@mathgroup\mv@normal\upmath@group{eur}{m}{n}
      \define@mathgroup\mv@bold\upmath@group{eur}{b}{n}
      \edef\UPM{\hexnumber\upmath@group}
      \new@mathgroup\amsa@group
      \define@mathgroup\mv@normal\amsa@group{msa}{m}{n}
      \define@mathgroup\mv@bold\amsa@group{msa}{m}{n}
      \edef\AMSa{\hexnumber\amsa@group}
      \makeatother
      \mathchardef\upi="0\UPM19
      \mathchardef\umu="0\UPM16
      \mathchardef\upartial="0\UPM40
      \mathchardef\leqslant="3\AMSa36
      \mathchardef\geqslant="3\AMSa3E
    \fi
  \fi
\fi % End of NFSS release 1

\ifnfsstwo
  \DeclareMathAlphabet{\mathbfit}{OT1}{cmr}{bx}{it}
  \SetMathAlphabet\mathbfit{bold}{OT1}{cmr}{bx}{it}
  \DeclareMathAlphabet{\mathbfss}{OT1}{cmss}{bx}{n}
  \SetMathAlphabet\mathbfss{bold}{OT1}{cmss}{bx}{n}
  \ifAMStwofonts
    \ifCUPmtlplainloaded \else
      \DeclareSymbolFont{UPM}{U}{eur}{m}{n}
      \SetSymbolFont{UPM}{bold}{U}{eur}{b}{n}
      \DeclareSymbolFont{AMSa}{U}{msa}{m}{n}
      \DeclareMathSymbol{\upi}{0}{UPM}{"19}
      \DeclareMathSymbol{\umu}{0}{UPM}{"16}
      \DeclareMathSymbol{\upartial}{0}{UPM}{"40}
      \DeclareMathSymbol{\leqslant}{3}{AMSa}{"36}
      \DeclareMathSymbol{\geqslant}{3}{AMSa}{"3E}
    \fi
  \fi
\fi % End of NFSS release 2

\ifCUPmtlplainloaded \else
  \ifAMStwofonts \else % If no AMS fonts
    \def\upi{\pi}
    \def\umu{\mu}
    \def\upartial{\partial}
  \fi
\fi

\title
[Reflected Iron Line From Kerr Black Hole Accretion Disc]
{Reflected Iron Line From a Source Above a Kerr Black Hole Accretion Disc}

\author[Y.~Dabrowski and A.N.~Lasenby]
{Y.~Dabrowski\thanks{Email: youri@mrao.cam.ac.uk} and A.N.~Lasenby\\
Astrophysics Group,
Cavendish Laboratory,
Madingley Road,
Cambridge,
CB3 0HE,
UK \\
}
\date{Accepted ???. Received ???; in original form \today}
\pagerange{\pageref{firstpage}--\pageref{lastpage}}
\pubyear{1997}

\begin{document}
\newcommand {\modela}{{\it this work (1)}}
\newcommand {\modelb}{{\it this work (2)}}
\newcommand {\modelc}{{\it this work (3)}}
\newcommand {\arccosh}{{\rm arccosh}}
\newcommand {\EW}{$W_{\rm K\alpha}$ }
\def\pdiff#1#2{{\frac{\partial #1}{\partial #2}}}
\def\Ddiff#1#2{{\frac{{\rm d} #1}{{\rm d} #2}}}

\newcommand {\observer}{$\gamma_0$-observer }
\newcommand {\observerc}{$\gamma_0$-observer}

\maketitle
\label{firstpage}

\begin{abstract}
In this paper we present a fully relativistic approach to modelling
both the continuum emission and the reflected fluorescent iron line
from a primary X-ray source near a Kerr black hole. The X-ray source
is located above an accretion disc orbiting around the black hole.
The source is assumed to be a static point source located on an
arbitrary position above the disc, on or off the axis of rotation.  We
carry out Monte Carlo simulations in order to estimate the iron line
spectrum as well as its equivalent width.  Because of the
gravitational lensing effect, an enhancement of the iron line is
expected when the primary source is located close to the central black
hole.  We find that for a source located on the axis of rotation the
enhancement is relatively modest. An observer at inclination
$30$~degrees would measure an equivalent width of $\sim$300~eV in the
extreme case of a maximally rotating black hole and a source located
at height $1.5$ gravitational radius from the centre. This corresponds
to an equivalent width enhancement factor of about 2 compared to the {\it
classical} value where no lensing effect comes into play.  However,
when allowing the source to be located off the axis of rotation, much
stronger enhancement can be obtained.  In the extreme case of a
maximally rotating black hole and a source located just above the
approaching side of the disc, an observer at inclination 30~degrees
could measure an equivalent width as high as $\sim$1.5~keV (i.e.
$\sim$10 times the classical value).  We also
find that observers located at high inclination angles observe a
stronger line than observers at low inclination angles.

\end{abstract}


\section{Introduction}

It is now commonly agreed that Active Galactic Nuclei (AGN) are
composed of a central supermassive black hole closely surrounded by a
thin equatorial accretion disc, the emission process being controlled
by viscous transport of angular momentum (see e.g. Rees 1984;
Blandford and Rees 1992).  The disc material is relatively {\it cold}
($\sim 10^6$~K) compared to the hard X-ray corona ($\sim 10^8$~K) by
which it is surrounded.  The corona is responsible for the observed
X-ray continuum emission which follows a power-law of index
$\Gamma\cong-1.7$. X-ray observation of Seyfert-1 galaxies revealed
that there exist spectral features and other deviations from a simple
power-law (e.g. Pounds et al. 1989; Nandra et al. 1989; Matsuoka et
al. 1990).  In particular a strong iron line is observed at 6.4~keV
together with an excess in the continuum emission at energies higher
than $\sim$4~keV.  This is well explained by Compton reflection
re-processing of the hard X-ray power-law continuum onto the accretion
disc (see e.g George \& Fabian 1991; Matt, Perola \& Piro 1991 ).  The
observed iron line is the result of fluorescent processes and the
excess in the continuum is accounted for by the presence of a
continuum component in the reflected emission.  The detailed spectrum
of the iron line is of great importance since it allows one to probe
regions of the accretion disc as close as a few Schwarzschild radii
from the central black hole (see e.g. Tanaka et al. 1995). In the
particular case of MCG--6-30-15, the lineshape yields evidence that
the emission occurs as close as half a Schwarzschild radius, giving
strong support for the presence of a rotating Kerr black hole (Tanaka
et al. 1995, Dabrowski et al. 1997).

The fluorescent iron line profile is well predicted by models which
assume a power-law $\epsilon(r)\propto r^{-\alpha}$ for the
fluorescent emissivity on the accretion disc, where $\alpha\sim 2$
(see e.g. Tanaka et al. 1995; Bromley, Chen \& Miller 1997; Dabrowski
et al. 1997; Reynolds \& Begelman 1997; Cadez et al. 1998). In such
cases the hard X-ray corona is assumed to be an extended source
located above and below the disc so that approximately half of the
emitted power illuminates the disc and is responsible for the
reflected continuum component as well as observed fluorescent lines
(George \& Fabian 1991). The remaining half escapes to infinity and
constitutes the major part of the observed continuum spectra.  The
predicted iron line equivalent width (EW) is in this case estimated to
be $\sim$100 - 200~eV (e.g. George \& Fabian 1991, Matt et al. 1991).
However, as discussed in Nandra et al. (1997), for a sample of 18
Seyfert-1 galaxies, observations reveal equivalent widths of 300-600
eV, or even as high as 1~keV in the case of MCG--6-30-15.  Part of the
EW enhancement may be explained by an overabundance of iron in the
accretion disc material (e.g. Lee et al. 1998, 1999).  However, other
means of enhancement need to be found in order to account for large
values of the iron line EW (Reynolds \& Fabian 1997).  Following the
work of Martocchia \& Matt (1996) and very recently Martocchia, Karas
\& Matt (1999), in this paper we propose to investigate the possible
EW enhancement due to the gravitational lensing of the primary light
rays which illuminate the disc and therefore drive the fluorescent
process.  The primary hard X-ray source is assumed to be a {\it
static} point source located above the accretion disc.  Situations
where the source is located on or off the axis of rotation are both
considered here.

Our predictions are based upon Monte Carlo simulations which account
for the primary X-ray emission as well the reflected iron line in a
consistent manner.  It is assumed here that the primary continuum
emission dominates over the reflected continuum emission so that the
reflected continuum is neglected while calculating the line EW (see
Section~\ref{sec:off}).  Details of our fully relativistic theoretical
approach are given in Section~\ref{sec:model}.  Some results are given
in the form of iron line profiles and equivalent widths in
Section~\ref{sec:on} for a primary source located on the axis of
rotation.  In Section~\ref{sec:off}, this assumption is relaxed and we
investigate line profiles and equivalent widths for arbitrary
positions of the source above the disc.  Finally our results are
discussed in Section~\ref{sec:discussion}.

\section{Reflection Model In the Kerr Metric}
\label{sec:model}
The black holes considered here belong to the Kerr family of solutions to
Einstein's equation and the metric employed is in the Boyer-Lindquist
form, which is defined by
\begin{eqnarray}
\label{eq:boyer-lindquist}
ds^2&=&dt^2-\rho^2\left(\frac{dr^2}{\Delta}+d\theta^2\right)-\left(r^2+L^2\right)\sin^2\theta d\phi^2 \nonumber \\
&&-\frac{2Mr}{\rho^2}\left(L\sin^2\theta d\phi-dt\right)^2,
\end{eqnarray}
where
\begin{equation}
\rho^2 = r^2+L^2\cos^2\theta,
\end{equation}
\begin{equation}
\Delta = r^2-2Mr+L^2.
\end{equation}
Here $t$, $r$, $\theta$ and $\phi$ are the space-time coordinates, $M$ is
the geometric mass, the mass of the black hole being $Mc^2/G$ and the
quantity $LM$ is the angular momentum of the hole as measured at
infinity (e.g. D'Inverno 1992).  The gravity of the disc itself is
assumed to be negligible and therefore the whole spacetime is
described by the Boyer-Lindquist metric.  We employ natural units
$G=c=\hbar=1$, unless stated otherwise.

\subsection{Photon Path}
\label{sec:photon}
Throughout this paper we use the gauge-theoretic formalism of Lasenby,
Doran \& Gull (1998) which allows fully relativistic and covariant
calculations to be made in the context of an easy-to-handle flat
space. It is important to note that predictions for measurable
quantities agree {\it exactly} with those of General Relativity for
this case.  Equations relevant to an equivalent approach to the
problem, but following a more standard general relativistic formalism,
are presented in Appendix~\ref{appendix:gr}.

The gauge-theoretic approach to gravity of Lasenby et al. (1998)
employs the language of {\it geometric algebra}, which seems to most
clearly expose the physics involved. In particular, the geometric
algebra of spacetime (Hestenes 1966), known as the Space Time Algebra
(STA), is generated by a set of four orthonormal vectors
\{$\gamma_{\mu}$\}, $\mu = 0...3$, satisfying
\begin{equation}
\label{eq:gammas}
\gamma_{\mu}\cdot\gamma_{\nu}=\eta_{\mu\nu}={\rm diag}(+---).
\end{equation}

In order to investigate problems in relation with the Kerr geometry,
we also introduce a suitable set of coordinates $t$, $r$, $\theta$ and
$\phi$ associated with the polar coordinate system and defined in
terms of the fixed \{$\gamma_{\mu}$\} frame. The corresponding
coordinate frame $e_{\mu}\equiv \partial_{\mu}x$ is
\begin{eqnarray}
\label{eq:e_t}
e_t & \equiv & \gamma_0\\
e_r & \equiv & \sin\theta\cos\phi\gamma_1 + \sin\theta\sin\phi\gamma_2 +
	       \cos\theta\gamma_3\\
e_{\theta} & \equiv & r\left( \cos\theta\cos\phi\gamma_1 +
                      \cos\theta\sin\phi\gamma_2 -
	              \sin\theta\gamma_3\right)\\
\label{eq:e_phi}
e_{\phi} & \equiv & r\sin\theta\left(-\sin\phi\gamma_1 + \cos\phi\gamma_2\right).
\label{def_e}
\end{eqnarray}
Note that $e_{\theta}$ and $e_{\phi}$ are not unit vectors and differ
by factors $r$ and $r\sin\theta$ from the usual spherical-polar
basis vectors, respectively. This is a direct consequence of their
definition in terms of the \{$\gamma_{\mu}$\} frame (see Lasenby et
al. 1998).

The translation and rotation gauge fields corresponding to the
Boyer-Lindquist metric (\ref{eq:boyer-lindquist}) are given in Doran,
Lasenby \& Gull (1996).

\begin{figure}
\centerline{\epsfig{file=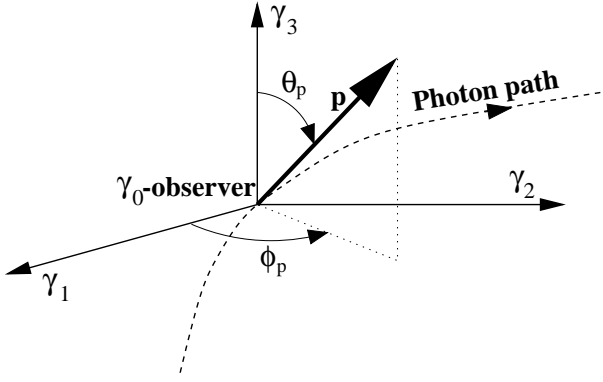, width=8cm}}
\caption{Photon 4-momentum $p$ as measured by the \observerc.
$\theta_p$ and $\phi_p$ are the observed angles of the photon
trajectory.}
\label{fig:p_angle_def}
\end{figure}

We start by defining an observer with covariant 4-velocity
equal to $\gamma_0$. Hereafter this observer will be called the
\observerc. One can show that, at infinity, the \observer is in
a flat Minkowski space-time and at rest with respect to the central
black hole.  Elsewhere its function is to provide a useful frame in
which to express quantities.  In order to parameterise the energy and
direction of the photon path, let us define the photon 4-momentum $p$
in the \{$\gamma_{\mu}$\} frame by
\begin{eqnarray}
\label{eq:p_def}
p&=&\Phi\gamma_0+\Phi\sin\theta_p\cos\phi_p\gamma_1+\Phi\sin\theta_p\sin\phi_p\gamma_2+\nonumber\\
&&\Phi\cos\theta_p\gamma_3.
\end{eqnarray}
The photon energy as measured by the \observer is given by
$p\cdot\gamma_0$ which, from equation (\ref{eq:p_def}), is equal to
$\Phi$. Similarly, as illustrated in Fig.~\ref{fig:p_angle_def}, the
angles $\theta_p$ and $\phi_p$ correspond to the usual spherical-polar
angles in the local spatial frame of the \observer and define the
direction at which photons are received (or emitted) by the
\observerc.
We note that $p$ is guaranteed to be null since $p\cdot p=0$, which
is what is required for a massless particle.
%
%\begin{equation}
%p\cdot p=\Phi^2\left(-\sin^2\theta_p\cos^2\phi_p-\sin^2\theta_p\sin^2\phi_p-\cos^2\theta_p\right)=0.
%\end{equation}
The photon trajectory itself is parameterised by the affine parameter
$\lambda$. The spacetime position of the photon ($t$, $r$, $\theta$,
$\phi$) is defined by the position vector
\begin{equation}
\label{eq:x_def}
x=te_t+re_r,
\end{equation}
where $t$, $r$, $\theta$ and $\phi$ are scalar functions of $\lambda$.
The photon geodesic equations are given in
Appendix~\ref{appendix:geodesic}. The equations we obtain are all
first order ordinary differential equations for $t$, $r$, $\theta$,
$\phi$, $\Phi$, $\theta_p$ and $\phi_p$.  As mentioned in the
appendix, the attractive aspect of this approach is that the geodesic
equations have been obtained without making use of conserved
quantities along the photon path such as the energy, angular momentum
or Carter constant.  Instead these quantities can be used {\it a
posteriori} as an effective check of both analytical and numerical
results (see Appendix~\ref{appendix:geodesic}).

\subsection{Accretion Disc}
\label{sec:disc}
We now employ a similar method as in Section~\ref{sec:photon} to study
the properties of an accretion disc rotating around a Kerr black hole.
In particular we are interested in the velocity and energy of the 
accreted material. The assumptions are the following:

(i) The accretion disc is a thin disc and lies in the equatorial plane
($\theta = \pi/2$), perpendicular to the axis of rotation of the
central Kerr black hole.

(ii) The orbits are stable circular orbits.  In the same way as we
defined the photon momentum (\ref{eq:p_def}), we start here by
defining the 4-velocity vector $v_d$ of a particle in circular orbit
of radius $r$ around the black hole. Since the motion occurs in the
equatorial plane, we can parameterise $v_d$ as a function of
$\gamma_0$, $\gamma_1$ and $\gamma_2$ only:
\begin{equation}
v_d=\cosh U\gamma_0 + \left(\sinh U\cos\omega \tau\right)\gamma_1
+\left(\sinh U\sin\omega \tau\right)\gamma_2,
\end{equation}
where $\omega$ represents the angular velocity and $\tau$ is the
proper time along the world-line of the particle in circular orbit.
$U$ is a function of $r$, $L$ and $M$, but not a function of $\tau$
since the orbit is circular. The parameterisation in $\cosh$ and
$\sinh$ has been chosen to ensure that
\begin{equation}
v_d\cdot v_d = \cosh^2U-\sinh^2U=1,
\end{equation}
as required for a massive particle. In the \{$e_{\mu}$\} frame, the
velocity takes the simple form:
\begin{equation}
\label{eq:v_def}
v_d=\cosh Ue_t + \frac{\sinh U}{r}e_{\phi}.
\end{equation}

\begin{figure}
\centerline{\epsfig{file=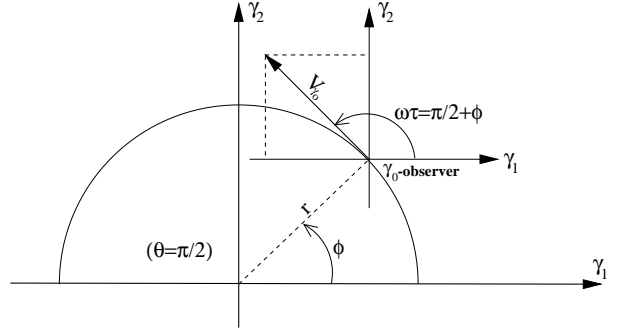, width=8cm}}
\caption{The 4-velocity of the orbiting particle, as measured by the
\observer at position ($r$, $\theta=\pi/2$, $\phi$). $\omega\tau=\pi/2+\phi$
and $V_{\gamma_0}=\tanh U$ are respectively the measured angle and
magnitude of the velocity.}
\label{fig:vel_def}
\end{figure}

We note that the \observer is not comoving with the particle in
orbit. Let us however assume that, at a given time $t$, the \observer
is at the position ($r$, $\theta=\pi/2$, $\phi$), i.e. instantaneously
at the same position as the orbiting particle (see
Fig.~\ref{fig:vel_def}).
According to the \observerc, the energy per unit mass of the particle
is given by $E_{\gamma_0} = v_d\cdot \gamma_0 = \cosh U$. The particle
velocity has a magnitude $V_{\gamma_0}=\tanh U$ and makes an angle
$\omega\tau=\pi/2+\phi$ from the $\gamma_1$ axis in the local spatial
frame ($\gamma_1$, $\gamma_2$, $\gamma_3$) of the \observerc.  When
solving the dynamical equations, we find
\begin{equation}
\label{eq:tanhU}
\tanh U = \frac{-L\pm\sqrt{Mr}}{\Delta^{\frac{1}{2}}},
\end{equation}
and therefore
\begin{eqnarray}
\cosh U &=& \frac{\Delta^{\frac{1}{2}}}{\sqrt{r^2-3Mr\pm2L\sqrt{Mr}}},\nonumber\\
\label{eq:coshUsinhU}
\sinh U &=& \frac{-L\pm\sqrt{Mr}}{\sqrt{r^2-3Mr\pm2L\sqrt{Mr}}}.
\end{eqnarray}
%
%As mentioned earlier, $\tanh U$ and $\cosh U$ are the tangential
%velocity $V_{\gamma_0}$ and energy per unit mass $E_{\gamma_0}$ of the
%orbiting particle as measured by a local \observerc, respectively.
In the case where $L>0$, the $+$ sign in (\ref{eq:tanhU}) and
(\ref{eq:coshUsinhU}) is for orbits co-rotating with the spinning
hole, while the $-$ sign is for counter-rotating orbits. If $L<0$ the
significance of the signs is reversed.

Finally, following the same approach as in
Appendix~\ref{appendix:geodesic}, the conserved energy per unit mass
$E_r$ and angular momentum $J_r$ of a circular orbit can be evaluated
by using the Killing vectors $K_t$ (equation \ref{eq:K_t}) and
$K_{\phi}$ (equation \ref{eq:K_phi}). We find
\begin{equation}
\label{eq:E_r}
E_r=K_t\cdot v_d=E_{\gamma_0}\frac{r^2-2Mr\pm L\sqrt{Mr}}{r\Delta^{\frac{1}{2}}}
\end{equation}
and
\begin{equation}
\label{eq:J_r}
J_r=K_{\phi}\cdot v_d = E_{\gamma_0}\frac{L\Delta^{\frac{1}{2}}+L^2V_{\gamma_0}}{r}+E_{\gamma_0}V_{\gamma_0}r.
\end{equation}
The conserved quantity $E_r$ in (\ref{eq:E_r}) is related to the local
energy $E_{\gamma_0}$ by a redshift factor, which tends to unity at
infinity. Therefore $E_r$ is usually called the {\it energy at
infinity}. Regarding $J_r$ in (\ref{eq:J_r}), the first term is a
purely gravitational effect due to the dragging effect of the spinning
hole, while the second term is more directly related to the Newtonian
expression, $mVr$, of the angular momentum. We note that the condition for
stability is given by $dE_r/dr > 0$, which gives the following condition for
the minimum stable orbit $r_{\rm ms}$:
\begin{equation}
\tanh U(r_{\rm ms}) = \frac{1}{2}.
\end{equation}
It is perhaps somewhat surprising that the minimum stable orbit is for
a local velocity of one half the speed of light. In this work, the
accretion disc extends from $r=r_{\rm in}=r_{\rm ms}$ up to the
arbitrary large outer radius $r=r_{\rm out}=5000~r_g$, where
$r_g=GM/c^2$ is the gravitational radius.

%The photon 4-momentum (\ref{eq:p_def}) is well defined for the
%\observer however, as mentioned earlier, this observer is not rotating
%along with the orbiting material.  It is therefore necessary to
%establish a relation between the 4-momentum energy $\Phi$ and
%direction ($\theta_p$, $\phi_p$) measured by the \observer and the
%equivalent quantities as measured by an observer co-rotating with the
%disc.  We first need to define a set of four orthogonall unit vectors
%with the time-like vector being the 4-velocity $v$ (\ref{eq:v_def}) of
%the orbiting particle. The orbits

\subsection{Primary Source and Distant Observer}
\begin{figure}
\centerline{\epsfig{file=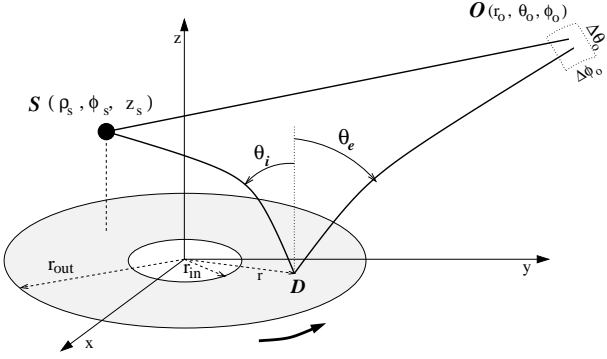, width=8cm}}
\caption{
Geometrical set up of our Monte Carlo simulations. The spinning black
hole is located at the centre of coordinates and the accretion disc
lies on the equatorial plane from the inner radius $r_{\rm in}$ to the
outer radius $r_{\rm out}$. The arrow indicates the sense of
rotation. The primary source is situated above the disc at an
arbitrary position ($\rho_s$, $\phi_s$, $z_s$) and illuminates both
the disc and the collecting area of a distant observer located at
($r_o$, $\theta_o$, $\phi_o$). The surface of the collecting area is
$r_o^2\sin\theta_o\Delta\phi_o\Delta\theta_o$.}
\label{fig:geo}
\end{figure}
The primary X-ray source is assumed to be a {\it static} point source
and, as shown in Fig.~\ref{fig:geo}, is located above the disc at an
arbitrary position ($\rho_s$, $\phi_s$, $z_s$), where $\rho$, $\phi$
and $z$ are the usual cylindrical polar coordinates.  The 4-velocity
$v_s$ of the source has to be proportional to the Killing vector $K_t$
(equation \ref{eq:K_t}) so that the source can remain static (see
e.g. Misner et al. 1973). Requiring that $v_s\cdot v_s=1$, we find
\begin{equation}
\label{eq:vs_def}
%v_s = \frac{\sqrt{\Delta-L^2\sin\theta_s}}{\rho}  K_t
v_s = \frac{1}{\sqrt{1-\frac{2Mr}{\rho^2}}}K_t
\end{equation}
The source is emitting isotropically in its proper frame with a rate
of emission defined by
\begin{equation}
N_{\nu_s} \propto \nu_s^{\Gamma}~{\rm photon~s^{-1}~Hz^{-1}},
\end{equation}
where $\Gamma$ is the photon power-law index. Throughout this work we
assume $\Gamma=-1.7$ (Mushotzky 1982; Turner \& Pounds 1989).

Predictions regarding the iron line flux and equivalent width are
those measured by a distant static observer.  Since $K_t=\gamma_0$ for
large $r$, the 4-velocity of the distant observer is
\begin{equation}
v_o=\gamma_0.
\end{equation}
As illustrated in Fig.~\ref{fig:geo}, the position of the distant
observer is given by $r_o$, $\theta_o$, $\phi_o$ with collecting
area on the $r=r_o$ sphere equal to
\begin{equation}
\Delta A=r_o^2\sin\theta_o\Delta\phi_o\Delta\theta_o.
\end{equation}

\subsection{Monte Carlo Simulations}

The Monte Carlo simulations we carry out in this work have the
advantage of computing both the continuum emission and the reflected
iron line flux as measured by a distant observer. In this manner the
equivalent width of the line can be estimated in a consistent manner.
A simulation consists of sending $N$ photons from the source
isotropically with respect to its proper frame. $N_c$ of these photons
will reach directly the observer's collecting area and therefore
contribute to the continuum flux. $N_l$ of them will be received after
having hit the disc and being reprocessed into iron line
photons. These contribute to the observed flux of the line. For each
photon received on the collecting area, the total spectrum is computed
by considering the total power in a succession of frequency bins of
extent $\Delta\nu$. The following two sections (\ref{sec:continuum} and
\ref{sec:line}) describe how the total power in each bin is
estimated, first for the continuum emission then the reflected
emission.

\subsubsection{Continuum Emission}
\label{sec:continuum}
Let us consider a photon emitted by the source with 4-momentum $p_s$
(equation \ref{eq:p_def}) which reaches directly the collecting area
at radius $r_o$ with 4-momentum $p_o$. The redshift measured by the
distant observer is therefore
\begin{equation}
1+z_{so} = \frac{p_s\cdot v_s}{p_o\cdot v_o},
\end{equation}
Since $v_o = \gamma_0$ the denominator is simply the value of $\Phi$
at the observer, $\Phi_o$.  The numerator is calculated using
(\ref{eq:vs_def}) and (\ref{eq:p_def}).  We find
\begin{equation}
\label{eq:1_zso}
1+z_{so} = \frac{\Phi_s\left[\Delta^{\frac{1}{2}}-L\sin\theta\sin\theta_p\sin(\phi_p-\phi)\right]_s}{\Phi_o},
\end{equation}
where the subscript $s$ indicates that the expression is evaluated at the
source.

We define $W^c_{\nu}$ to be the continuum total power received in the
collecting area of the distant observer in the range of frequency
[$\nu$, $\nu+\Delta\nu$]. We have
\begin{equation}
\label{eq:wc}
W^c_{\nu} = \sum_{N_c}{\int_{\nu}^{\nu+\Delta\nu}W_{\nu_o}{\rm d}\nu_o},
\end{equation}
where the sum is for each received continuum photon, $W_{\nu_o}$ is
the received power per unit frequency attached to the photon, and the
subscript $o$ denotes quantities as measured by the distant observer.
Expressing quantities on the right hand side of equation~(\ref{eq:wc})
in terms of quantities defined at the source, we obtain
\begin{equation}
W^c_{\nu} =
\sum_{N_c}\left(\frac{1}{1+z_{so}}\right)^2\int_{\nu(1+z_{so})}^{(\nu+\Delta\nu)(1+z_{so})}\nu_s^{\Gamma+1}{\rm
d}\nu_s.
\end{equation}

\subsubsection{Reflected Iron Line}
\label{sec:line}

We now consider a photon that leaves the source with a 4-momentum
$p_s$ and hits the disc with a 4-momentum $p_d$. The redshift measured
by the co-rotating observer is
\begin{equation}
1+z_{sd} = \frac{p_s\cdot v_s}{p_d\cdot v_d}.
\end{equation}
The numerator is given in (\ref{eq:1_zso}) and the denominator can be
calculated using (\ref{eq:p_def}) and (\ref{eq:v_def}). We find
\begin{equation}
1+z_{sd}= \frac
{\Phi_s\left[\Delta^{\frac{1}{2}}-L\sin\theta\sin\theta_p\sin(\phi_p-\phi)\right]_s}
{\Phi_d\Big[\cosh U-\sinh U\sin\theta_p\sin(\phi_p-\phi)\Big]_d},
\end{equation}
where the subscripts $s$ and $d$ indicate that the expression is evaluated
at the source and at the disc, respectively. A similar approach can be
followed for the second part of the trajectory, from the disc to the
collecting area of the distant observer. We find
\begin{equation}
1+z_{do} = \frac{\Phi_d\Big[\cosh U-\sinh U\sin\theta_p\sin(\phi_p-\phi)\Big]_d}{\Phi_o}.
\end{equation}

The radiation processes occurring on the disc are described in detail
in George \& Fabian (1991). Here, we are simplifying the situation
significantly by ignoring the reflected continuum emission so that
only the reflected iron line is taken into account (see
Section~\ref{sec:discussion} for a discussion of the validity of this
approach).  We define $N^{in}_{\nu_d}(\nu_d, \theta_i)$ to be the
total number of photons impinging on the disc with energy $\nu_d$ and
incidence angle $\theta_i$ per unit frequency and per unit time, as
measured by a co-rotating observer (see Fig.~\ref{fig:geo}).
Following George \& Fabian (1991), the number of fluorescent photons
per unit frequency and per unit time which are able to escape the disc
is
\begin{equation}
N^{out}_{\nu_d}= N^{in}_{\nu_d}(\nu_d, \theta_i)g(\theta_i)f(\nu_d),
\end{equation}
where,
\begin{eqnarray}
g(\theta_i) &=& \left(6.5-5.6\cos\theta_i+2.2\cos^2\theta_i\right)\times 10^{-2},\\
f(\nu_d) &=& 7.4\times 10^{-2}+2.5\exp{\left(-\frac{\nu_d-1.8}{5.7}\right)}.
\end{eqnarray}
This analytical approximation is valid for $\nu_t<\nu_d<\nu_m$, where
$\nu_t=7.1~{\rm keV}$ is the energy threshold for trigerring the
$6.4~{\rm keV}$ iron fluorescent emission, and $\nu_m=30~{\rm keV}$.
As given in Ghisellini, Haardt \& Matt (1994), the intensity of the
emerging fluorescent emission is not isotropic and is here assumed to
be proportional to $\cos\theta_e\ln(1+1/\cos\theta_e)$, where
$\theta_e$ is the outgoing inclination angle as measured in the
co-rotating frame. Therefore $N^{out}_{\nu_d}$ follows the angular
distribution
\begin{equation}
N^{out}_{\nu_d}(\theta_e){\rm d}\theta_e = 2\cos\theta_e\ln\left(1+\frac{1}{\cos\theta_e}\right)\sin\theta_e{\rm d}\theta_e.
\end{equation}

We now define $W^l_{\nu}$ to be the iron line total power received in
the collecting area of the distant observer in the range of frequency
[$\nu$, $\nu+\Delta\nu$]. Following a similar reasoning as in
Section~\ref{sec:continuum} we obtain
\begin{equation}
W^l_{\nu} = \sum_{N_l}{\delta_{\nu}\left(\frac{1}{1+z_{do}}\right)^2N^{out}\nu_{\alpha}},
\end{equation}
where $\nu_{\alpha}=6.4~{\rm keV}$, $\delta_{\nu}=1$ if
$\nu<\nu_{\alpha}/(1+z_{do})<\nu+\Delta\nu$, $\delta_{\nu}=0$
otherwise, and
\begin{equation}
N^{out} = g(\theta_i)\int_{\nu_t}^{\nu_m}f(\nu_d)N^{in}_{\nu_d}{\rm d}\nu_d.
\end{equation}
When expressed in the source proper frame, we have
\begin{eqnarray}
W^l_{\nu} = \sum_{N_l}\delta{\nu}\left(\frac{1}{1+z_{do}}\right)^2\frac{1}{1+z_{sd}}\nu_{\alpha}g(\theta_i)\times\nonumber\\
\int_{\nu_t(1+z_{sd})}^{\nu_m(1+z_{sd})}f\left(\frac{\nu_s}{1+z_{sd}}\right)
\nu_s^{\Gamma}{\rm d}\nu_s.
\end{eqnarray}

\section{Source on Axis}
\label{sec:on}
When discussing the fluorescent iron line profile and equivalent
width, many authors have considered the situation where the primary
X-ray source is located above the disc, on the axis of rotation
(e.g. Martocchia et al. 1999, Matt et al. 1992, Martocchia \& Matt
1996; Reynolds \& Begelman 1997, Young, Ross \& Fabian 1998; Reynolds
et al. 1999).  This assumption has the advantage of simplifying both
calculations and analysis since the axial symmetry of the problem is
conserved.  In this section we also follow this assumption, but it
will be relaxed in Section~\ref{sec:off}.

\subsection{Gravitational Lensing Effect}
\label{sec:lensing_on}

\begin{figure}
\centerline{\epsfig{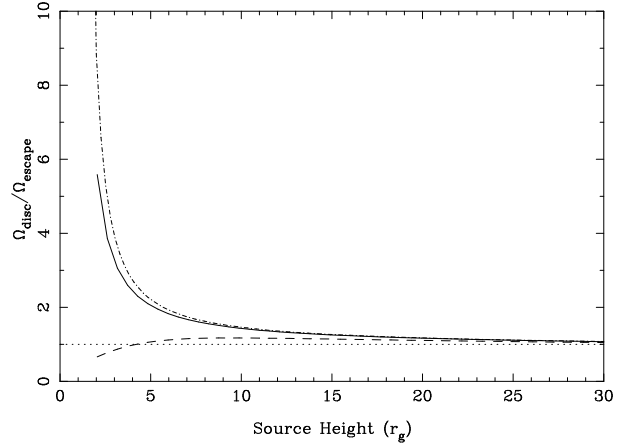}}
\caption{
Ratio of the solid angle formed by the rays hitting the disc to the
solid angle formed by rays escaping to infinity.  The solid line is
for a maximally rotating Kerr black hole while the dashed curve is for
a Schwarzschild hole.  The dotted-dashed curve is for the ratio of
solid angle formed by rays deflected towards the equatorial plane,
including those lensed towards the event horizon. The dotted line
indicates the ratio that would be obtained without the gravitational
bending effect.}
\label{fig:solid_angle}
\end{figure}
As pointed out by Martocchia \& Matt (1996), if the primary source is
very close to the hole, a significant fraction of the primary photon
rays will be lensed towards the disc, decreasing considerably the
direct flux relatively to the reflected flux, as observed by a distant
observer. In order to quantify this phenomena, the ratio of the solid
angle formed by the rays hitting the disc to the solid angle formed by
the rays escaping to infinity ($\Omega_{\rm disc}/\Omega_{\rm
escape}$) is plotted versus the height $h=z_s$ of the source in
Fig.~\ref{fig:solid_angle}.  The effect is negligible in the case of a
Schwarzschild black hole since a large amount of the lensed radiation
is lost within the innermost stable orbit of the disc
(i.e. 6~$r_g$). In the case of a maximally rotating Kerr black hole,
where the inner radius of the disc can be as small as 1~$r_g$, the
effect becomes significant for a sufficiently close source, typically
$h<6~r_g$ as seen on Fig.~\ref{fig:solid_angle}.  However, in the case
of a source at height $h=2~r_g$, $\Omega_{\rm disc}$ is only a factor
of $\sim 5.5$ bigger than $\Omega_{\rm escape}$. In this case we find
that almost half of the rays bent towards the equatorial plane
do not actually hit the disc but are strongly lensed towards the
event horizon, resulting in a significant damping of the effect.

\subsection{Line Profiles}
\label{sec:line_on}

\begin{figure}
\centerline{\epsfig{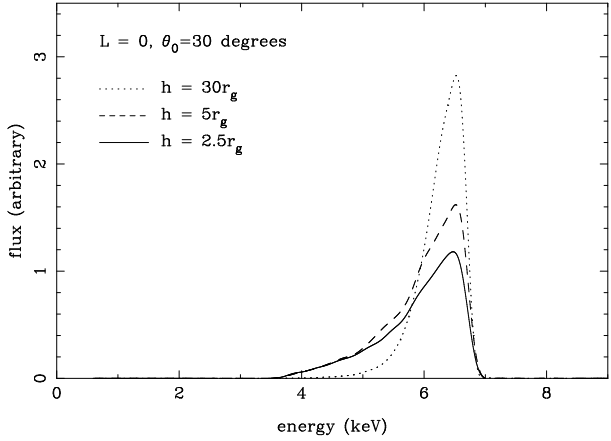}}
\caption{
Predicted spectral line shapes for a central Schwarzschild black hole
($L=0$), as seen by a distant observer at inclination angle
$\theta_0=30$~degrees.  The primary source is located on the axis of
rotation at height $h$ from the origin. The solid, dashed and dotted
lines are for $h=2.5$, $5$ and $30r_g$ respectively.  The area under
each curve is proportional to the line equivalent width.}
\label{fig:a0_on_30}
\end{figure}

\begin{figure}
\centerline{\epsfig{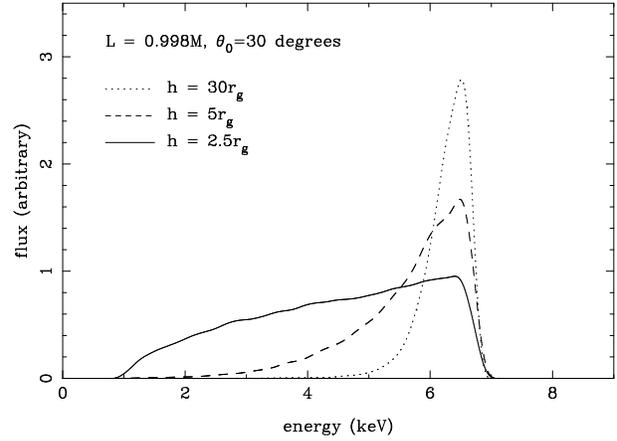}}
\caption{
Predicted spectral line shapes for a maximally rotating Kerr black hole
($L=0.998M$), as seen by a distant observer at inclination angle
$\theta_0=30$~degrees.  The primary source is located on the axis of
rotation at height $h$ from the origin. The solid, dashed and dotted
lines are for $h=1.5$, $5$ and $30r_g$ respectively.  The area under
each curve is proportional to the line equivalent width.}
\label{fig:kerr_on_30}
\end{figure}

\begin{figure}
\centerline{\epsfig{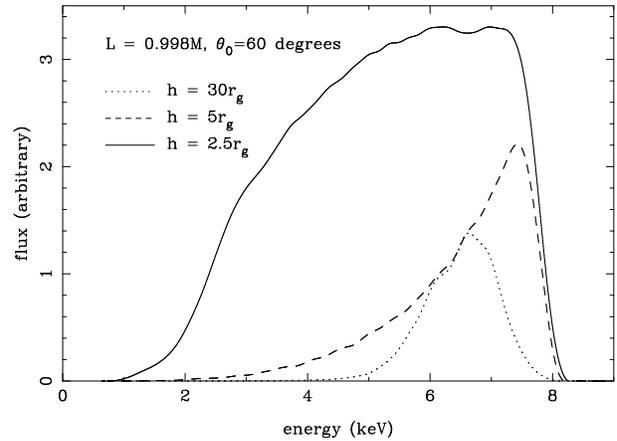}}
\caption{
Same as Fig.~\ref{fig:kerr_on_30} but for $\theta_0=60$~degrees.
}
\label{fig:kerr_on_60}
\end{figure}

Examples of line profiles obtained with our Monte-Carlo simulations
for various values of $h$ are given in Figs.~\ref{fig:a0_on_30},
\ref{fig:a0_on_60}, \ref{fig:kerr_on_30} and \ref{fig:kerr_on_60} for
the maximum Kerr and the Schwarzschild cases. The predicted line
shapes are for distant observers at inclination angles
$\theta_0=30$~degrees and $\theta_0=60$~degrees.
\begin{figure}
\centerline{\epsfig{file=fig6.eps, angle=-90, width=8cm}}
\caption{
Same as Fig.~\ref{fig:a0_on_30} but for $\theta_0=60$~degrees.
}
\label{fig:a0_on_60}
\end{figure}
As already noticed in previous studies (e.g. Fabian et al. 1989,
Dabrowski et al. 1997), the line extends further towards the blue
shifted part of the spectrum for larger inclination angles.  Indeed,
for large $\theta_0$ the blue part of the line is boosted by the
Doppler effect due to the rotation of the disc.  On the other hand,
the extension towards the red shifted part of the spectrum is
controlled by the strength of the gravitational redshift which
increases as the fluorescent emission happens closer to the central
black hole.  This is very clearly illustrated here for the maximally
rotating Kerr case (Figs~\ref{fig:kerr_on_30}, \ref{fig:kerr_on_60}),
where $\nu/\nu_{\alpha}$ could be as small as $0.16$ for a source
located at $h=1.5~r_g$. Indeed, as the source gets closer to the hole
it illuminates more intensively the inner part of the disc, resulting
in strong gravitational redshift of the fluorescent line.  In the
Schwarzschild case as well (Figs. \ref{fig:a0_on_30},
\ref{fig:a0_on_60}) the reddening of the line increases as the
source gets closer to the hole.  However, because the inner radius of
the disc is $6~r_g$ as opposed to $1.24~r_g$ in the maximally rotating
Kerr case, less strong reddening is predicted in the Schwarzschild
case. For example, one can notice very little difference between the
predicted lines for $h=5~r_g$ and $h=2.5~r_g$.  In particular the red
wing is not more extended in the $h=2.5~r_g$ case because most of the
illumination is focused within the minimum stable orbit.

\subsection{Equivalent Width}
\label{sec:ew_on}

\begin{figure}
\centerline{\epsfig{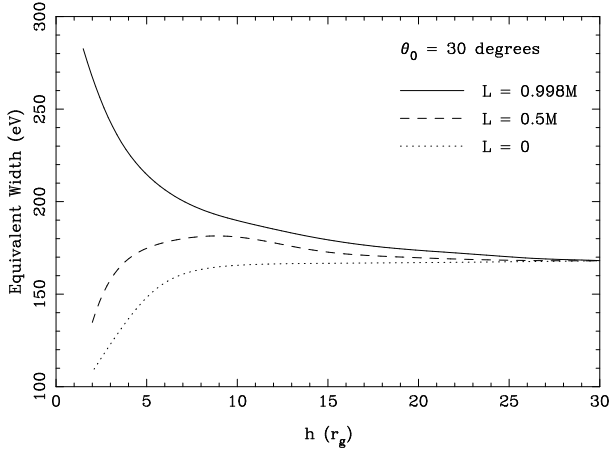}}
\caption{
Equivalent width of the iron fluorescent line as a function of the
height $h$ of the source. The solid, dashed, and dotted lines are for
$L=0.998M$, $L=0.5M$ and $L=0$ respectively. The distant observer is at
inclination $\theta_0=30$~degrees.
}
\label{fig:ew_on_30}
\end{figure}

\begin{figure}
\centerline{\epsfig{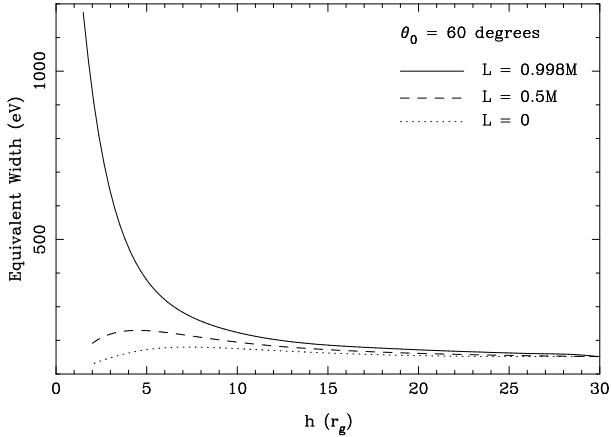}}
\caption{
Same as Fig.\ref{fig:ew_on_30} but for an observer at inclination
$\theta_0=60$~degrees.
}
\label{fig:ew_on_60}
\end{figure}

The equivalent width (EW) of the iron fluorescent line is estimated
as follows:
\begin{equation}
{\rm EW}=\frac{\int_0^\infty W^l_{\nu} {\rm d}\nu}{W^c_{\nu=\nu_{\alpha}}}.
\end{equation}
Results are given in Figs.~\ref{fig:ew_on_30} and \ref{fig:ew_on_60}
as a function of the height of the primary source in the maximally
rotating Kerr case ($L=0.998M$), in the Schwarzschild case ($L=0$) and
in an intermediate case where $L=0.5M$. Fig.~\ref{fig:ew_on_30} is for
a distant observer with inclination $\theta_0=30$~degrees and
Fig.~\ref{fig:ew_on_60} is for $\theta_0=60$~degrees.  
We expect the strength of the line to be larger as the source gets
closer to the hole since the primary emission would preferentially be
focused towards the plane of the disc. This effect is clearly visible
in the case of a maximally rotating Kerr black hole (solid lines in
Figs.\ref{fig:ew_on_30} \& \ref{fig:ew_on_60}).  However, for lower
values of the black hole spin and consequently higher values of the
inner radius of the disc, this EW enhancement effect is lessened, as
predicted in Section~\ref{sec:lensing_on}.  In cases where the source
is so close to the hole that most of the lensed illumination is lost
within the inner radius of the disc, the effect is actually reversed
and the EW diminishes as the primary source gets closer to the hole.
In particular, no EW enhancement is predicted in the case of a
Schwarzschild black hole.  For a distant observer at inclination
$\theta_o=30$~degrees the maximum EW is predicted to be $\sim$290~eV,
while the EW obtained for the {\it classical} case where no lensing
effect comes into play (i.e. large $h$) is $\sim$165~eV.  Therefore,
even for a primary source located just above the event horizon of a
maximally rotating Kerr black hole, the EW enhancement is less than a
factor of two.  As seen in Fig.\ref{fig:ew_on_60}, the gravitational
lensing effect becomes more effective when observed from a large
inclination angle.  Indeed, the line is strengthened by the lensing
effect for both low and high inclination angle points of view.
However, due to the same effect, the continuum emission received at
large $\theta_o$ is strongly reduced compared to that received at
smaller $\theta_o$.  For an observer at inclination
$\theta_o=60$~degrees, EWs as high as $\sim$1.2~keV are predicted,
which represents an EW enhancement factor of about 8.  This strong EW
enhancement effect can also be seen in Fig.~\ref{fig:kerr_on_60},
where the area under each line is proportional to the line EW.
However in the case of MCG--6-30-15 the disc is thought to have a
inclination angle of about 30 degrees (e.g. Tanaka et al.  1995;
Dabrowski et al. 1997).

\section{Source off Axis}
\label{sec:off}
We have seen in Section~\ref{sec:on} that the EW enhancement is limited
by the fact that a significant proportion of the primary illumination
is `lost' within the inner radius of the disc instead of contributing
to the strengthen of the iron line.  This limiting effect is
particularly important when the primary source lies above the disc on
the axis of rotation.  Stronger EW enhancement should be obtained by
relaxing this assumption and allowing the primary source not to lie
right above the event horizon of the hole.  In this section, we
investigate this possibility in the maximally rotating Kerr solution
($L=0.998M$).

\begin{figure}
\centerline{\epsfig{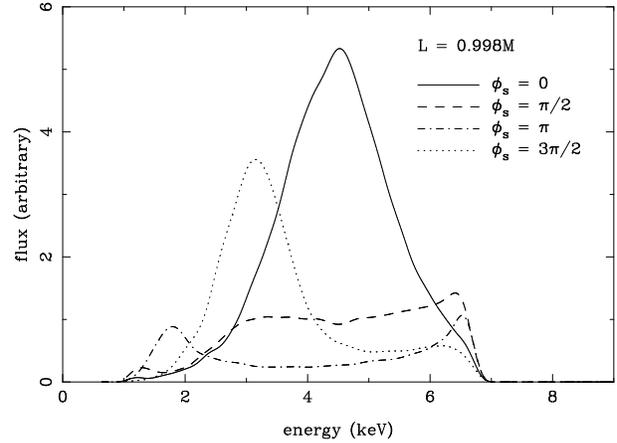}}
\caption{
Predicted spectral line shapes for a maximally rotating black hole
($L=0.998M$), as seen by a distant observer at inclination angle
$\theta_o=30$~degrees and azimuthal angle $\phi_o=\pi/2$. The primary
sources are located at $\rho_s=2.5~r_g$ and $z_s=1~r_g$, in each case.
The solid, dashed, dashed-dot and dotted lines are for $\phi_s=0$,
$\phi_s=\pi/2$, $\phi_s=\pi$ and $\phi_s=3\pi/2$ respectively. The
area under each curve is proportional to the line equivalent width.}
\label{fig:lines}
\end{figure}

Since the axial symmetry of the system is here broken, we need to
define the distant observer by both the inclination $\theta_o$ and the
azimuthal angle $\phi_o$. For simplicity, here we fix the observer at
$\phi_o=\pi/2$ while allowing the source to be located at various
azimuths $\phi_s$ above the disc, as suggested in Fig.~\ref{fig:geo}.
Line spectra are given in Fig.~\ref{fig:lines} as a function of
$\phi_s$ for sources located at height $z_s=1~r_g$ above the plane of
the disc and radius $\rho_s=2.5~r_g$ from the axis of rotation, close
to the ergosphere radius at this latitude ($1.93~r_g$).
Since the illumination takes place above the inner parts of the disc,
all these lines display strongly redshifted features. However, as one
can notice on Fig.~\ref{fig:lines}, the shape of the spectral line
varies significantly as a function of $\phi_s$.  For $\phi_s=0$ the
line is mostly redshifted and its flux peaks at around 4.5~keV, while
for $\phi_s=\pi$ the line is double peaked around 1.8~keV and 6.5~keV.
In the former case the primary source is located above the approaching
part of the disc where the emission is significantly boosted by the
Doppler effect. In the latter case, the strongly redshifted red wing
is the result of Doppler shift from the receding part of the disc,
while the blue wing is due to rays that have reached the approaching
side of the disc while spiralling down on to the disc.  Lines for
$\phi_s=\pi/2$ and $\phi_s=3\pi/2$ are hybrids of the above two cases.

\begin{figure}
\centerline{\epsfig{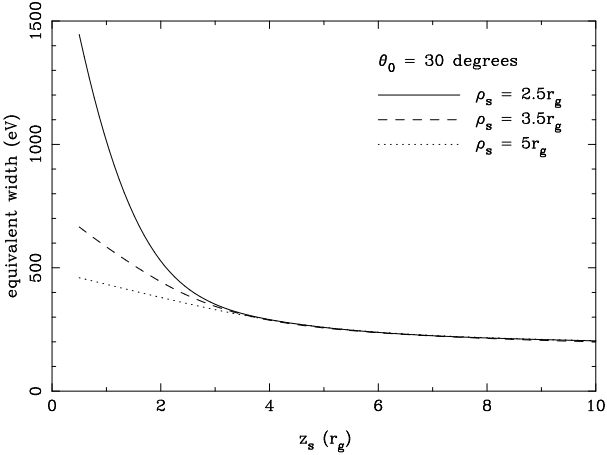}}
\caption{
Equivalent width of the iron fluorescent line as a function of the
height $z_s$ of the source. The primary source is at azimuth
$\phi_s=0$, above the approaching side of the disc. The solid, dashed
and dotted lines are for $\rho_s=2.5$, $3.5$ and $5~r_g$ respectively.
The central black hole is a maximally rotating black hole ($L=0.998M$) and
the distant observer is at inclination $\theta_o$=30~degrees and 
azimuth $\phi_o=\pi$.
}
\label{fig:ew_off_30}
\end{figure}

\begin{figure}
\centerline{\epsfig{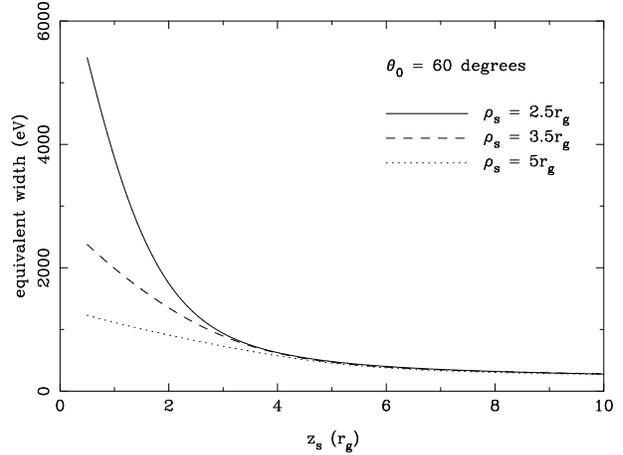}}
\caption{
Same as Fig.~\ref{fig:ew_off_30} but for an observer at inclination
$\theta_o$=60~degrees.
}
\label{fig:ew_off_60}
\end{figure}

Since the illumination occurs right above the Doppler boosted region
of the disc for $\phi_s=0$, the predicted line EW is much larger in
this geometrical situation than for other values of $\phi_s$.  This is
illustrated in Fig.~\ref{fig:lines} where the spectral line integrals
are proportional to their EW.  For the rest of this section we assume
that the primary source is always located at this privileged azimuthal
angle $\phi_s=0$. Figs.~\ref{fig:ew_off_30} and \ref{fig:ew_off_60}
give the predicted iron line EW as a function of the primary source
height $z_s$ and radius $\rho_s$.
In the case of an observer at inclination $\theta_o$=30~degrees, the
observed EW can be much larger than for a source located on the axis
of rotation. For example, as seen on Fig.\ref{fig:ew_off_30}, for a
primary source located at ($\rho_s=2.5~r_g$, $\phi_s=0$, $z_s=1~r_g$)
the predicted EW is $\sim$1~keV and if the source is as close as
$z_s=0.5~r_g$ to the disc the EW can be as high as $\sim$1.5~keV. In
this latter case, the EW is enhanced by a factor of $\sim$10 of its
{\it classical} value.  However, in order to reach such high EW the
primary source needs to be located above the very inner parts of the
disc. For example, in the case of a source located at ($\rho_s=5~r_g$,
$\phi_s=0$, $z_s=0.5~r_g$) the predicted line EW is only $\sim$500~eV.
As a consequence, very little EW enhancement is expected in the case
of slowly rotating black holes. As expected, for an observer at
inclination $\theta_o$=60~degrees the predicted line EW is much larger
than for $\theta_o$=30~degrees.  As seen on Fig.\ref{fig:ew_off_60},
in the extreme case where the primary source is located at
($\rho_s=2.5~r_g$, $\phi_s=0$, $z_s=0.5~r_g$), the observed EW can be
as high as $\sim$5.5~keV, which represents an EW enhancement factor of
about 35.  

\section{Discussion}
\label{sec:discussion}
In Section~\ref{sec:model} we have presented a new fully relativistic
approach to calculating photon paths and energies in the
Boyer-Lindquist metric.  The geodesic equations obtained in this
approach are simple first order ordinary differential equations which
allow easy and stable numerical integration.  Furthermore, the way the
photon 4-momentum is parameterised (equation \ref{eq:p_def}) allows us
to be very clear regarding observable quantities such as the frequency
and incoming/outgoing angles of the light ray, as measured by an
observer.  We believe that our theoretical approach is an
improvement compared to more standard methods usually employed in
similar applications (e.g. Reynolds et al. 1999).

We found in Sections~\ref{sec:on} and \ref{sec:off} that large values
of the observed iron line equivalent width (EW) may be explained by
general relativistic effects when the primary X-ray source lies above
the very inner part of an accretion disc (within $\sim$5 gravitational
radii) around a maximally rotating central black hole.  When the
primary source is located on the axis of rotation, EWs up to
$\sim$300~eV only can be accounted for by this model for a disc at
inclination 30 degrees.  This is a much lower prediction than that of
the early work of Martocchia \& Matt (1996) who found EWs up to
$\sim$1500~eV for a comparable case.  Our results are however in good
agreement with the very recent work of Martocchia et al. (1999).  At
inclination 60 degrees EWs up to $\sim$1.2~keV are predicted.

By allowing the primary source to be located off the axis of rotation, we
found that much larger values of the EW are
obtained when the source is located above the approaching side of the
disc (up to $\sim$1.5~keV and $\sim$5.5~keV for inclination 30 and 60 degrees 
respectively).  However, in the most extreme cases, a reflected continuum
component should also be included in the model, in addition to the
reflected iron line (Lightman \& White 1988; Pounds et al. 1990; Matt
et al. 1991; George \& Fabian 1991 ).  When the reflected emission is
not enhanced by the lensing of the primary light rays, the reflected
continuum accounts for about 10 percent of the observed continuum flux
at 6.4~keV. One would therefore expect the reflected continuum to
dominate when the enhancement factor reaches $\sim$10.  Beyond this
value, the continuum flux observed at 6.4~keV increases together with
the iron line flux and therefore the line should no longer gain in
strength.  As a result one should regard values of
Fig. \ref{fig:ew_off_60} as indicative of the strength of the lensing
enhancement effect, while the actual observed iron line EW should not
be larger than $\sim$~1600~eV (i.e. 10 times its {\it classical}
value).

Results found in Section~\ref{sec:off} may be used as arguments
towards explaining the large iron line EW observed in Seyfert-1
galaxies. Such an explanation may be valid for the EW enhancement
observed in the case of MCG--6-30-15 where it is likely that strong
emission processes take place very close to the hole, at least during
specific periods (see Iwasawa et al. 1996, 1999).  However, as
highlighted by Reynolds \& Fabian (1997), in the case of more typical
objects it is less probable that strong emission would occur so close
to the hole.  Reynolds \& Fabian (1997) found that the relative motion
between the disc and the source may significantly affect the EW of the
line, simply by special relativistic arguments.  This could allow EW
enhancement even when the emission occurs further away from the hole.
Such an effect could be investigated in our model by allowing the
source to have a peculiar velocity. For example, a simple model where
the source moves with a constant angular velocity could be assumed (Yu
\& Lu 1999).

\section*{Acknowledgements}

YD would like to thank Trinity Hall, Cambridge for support in the form
of a Research Fellowship. The authors also thank Andy Fabian and Anthony
Challinor for useful discussions.

\appendix

\section{General Relativistic Approach}
\label{appendix:gr}
This work has been carried out using the gauge-theoretic formalism
described in detail in Lasenby et al. (1998). However, the theoretical
calculations presented in Section~\ref{sec:model} and
Appendix~\ref{appendix:geodesic} could in principle be carried out
following a more standard general relativistic approach by making use
of orthogonal tetrads (see e.g. Misner et al. 1973).  A suitable
tetrad frame in the Boyer-Lindquist metric
(equation~\ref{eq:boyer-lindquist}) can be defined as follows:
\begin{eqnarray}
\left(e_t\right)^{\mu} &=& \bigg(\frac{r^2+L^2}{\rho\Delta^{\frac{1}{2}}},
	0, 0, \frac{L}{\rho\Delta^{\frac{1}{2}}}\bigg)\\
\left(e_r\right)^{\mu} &=& \bigg(0, \frac{\Delta^{\frac{1}{2}}}{\rho},
	0, 0\bigg)\\
\left(e_{\theta}\right)^{\mu} &=& \bigg(0,0,\frac{r}{\rho},0\bigg)\\
\left(e_{\phi}\right)^{\mu} &=& \bigg(\frac{Lr\sin^2\theta}{\rho},
	0, 0, \frac{r}{\rho}\bigg),
\end{eqnarray}
where $\mu = 0\ldots3$ are the usual coordinate indexes, corresponding
here to the Boyer-Lindquist coordinates ($t$, $r$, $\theta$, $\phi$).
This tetrad can be identified with the coordinate frame \{$e_{\mu}$\}
defined by equations~(\ref{eq:e_t}) to (\ref{eq:e_phi}).  The
orthonormal tetrad corresponding to the \{$\gamma_{\mu}$\} frame
(equation
\ref{eq:gammas}) is given by
\begin{eqnarray}
\left(\gamma_0\right)^{\mu} &=& \bigg(\frac{r^2+L^2}{\rho\Delta^{\frac{1}{2}}},
	0, 0, \frac{L}{\rho\Delta^{\frac{1}{2}}}\bigg)\\
\left(\gamma_1\right)^{\mu} &=&  \bigg(
	\frac{-L\sin\theta\sin\phi}{\rho},
	\frac{\Delta^{\frac{1}{2}}\sin\theta\cos\phi}{\rho},\nonumber \\
&&~~~
	\frac{\cos\theta\cos\phi}{\rho},
	\frac{-\sin\phi}{\rho\sin\theta}\bigg)\\
\left(\gamma_2\right)^{\mu} &=&  \bigg(
	\frac{L\sin\theta\cos\phi}{\rho},
	\frac{\Delta^{\frac{1}{2}}\sin\theta\sin\phi}{\rho},\nonumber \\
&&~~~
	\frac{\cos\theta\sin\phi}{\rho},
	\frac{\cos\phi}{\rho\sin\theta}\bigg)\\
\left(\gamma_3\right)^{\mu} &=& \bigg(0,
	\frac{\Delta^{\frac{1}{2}}\cos\theta}{\rho},
	\frac{-\sin\theta}{\rho}, 0\bigg).
\end{eqnarray}
The equations given in Section~\ref{sec:model} can now be
translated into this general relativistic context. For example
equation~(\ref{eq:p_def}) of Section~\ref{sec:photon} should be read
as follows
\begin{eqnarray}
p^{\mu} &=& \Phi\left(\gamma_0\right)^{\mu}+
	    \Phi\sin\theta_p\cos\phi_p\left(\gamma_1\right)^{\mu}+\nonumber\\
        & & \Phi\sin\theta_p\sin\phi_p\left(\gamma_2\right)^{\mu}+
	    \Phi\cos\theta_p\left(\gamma_3\right)^{\mu}.
\end{eqnarray}
While the photon geodesic equations $\dot{t}$, $\dot{r}$,
$\dot{\theta}$, $\dot{\phi}$ given in Appendix~\ref{appendix:geodesic}
are obtained by the following equation
\begin{equation}
p^{\mu}=\frac{\left({\rm d}x\right)^{\mu}}{{\rm d}\lambda},
\end{equation}
where 
\begin{equation}
\left({\rm d}x\right)^{\mu}=\left({\rm d}t, {\rm d}r, {\rm d}\theta, {\rm d}\phi\right).
\end{equation}

\onecolumn

%\begin{minipage}{155mm}
\section{Photon Geodesic Equations}
\label{appendix:geodesic}
When solving the geodesic equations in the Boyer-Lindquist metric for
the photon 4-momentum given in Section~\ref{sec:photon} (see Lasenby
et al. 1998), we find
\begin{eqnarray}
\label{eq:system1}
\frac{{\rm d}t}{{\rm d}\lambda}&=&
\frac {\Phi}{\rho\Delta^{\frac{1}{2}}}
    \Big[{L}^{2}+{r}^{2}+L\Delta^{\frac{1}{2}}\sin\theta\sin\theta_p
\sin(\phi_p-\phi) \Big],\nonumber\\  
\frac{{\rm d}r}{{\rm d}\lambda}&=&
\frac {\Phi\Delta^{\frac{1}{2}}}{\rho}
\Big[\cos\theta_p\cos\theta+
\sin\theta\sin\theta_p\cos(\phi_p-\phi) \Big],\\ 
\frac{{\rm d}\theta}{{\rm d}\lambda}&=&
\frac {\Phi}{\rho}
\Big[\cos\theta\sin\theta_p\cos(\phi_p-\phi)-
\sin\theta\cos\theta_p\Big],\nonumber\\
\frac{{\rm d}\phi}{{\rm d}\lambda}&=&
\frac {\Phi}{\rho\Delta^{\frac{1}{2}}\sin\theta}
\Big[\Delta^{\frac{1}{2}}\sin\theta_p\sin(\phi_p-\phi)+
L\sin\theta \Big],\nonumber
\end{eqnarray}
and for the photon 4-momentum
\begin{eqnarray}
\label{eq:system2}
\frac{{\rm d}\Phi}{{\rm d}\lambda}&=&
\frac{\Phi^2}{\rho^3}\Bigg\{
\Big[\cos\theta\cos\theta_p+\sin\theta\sin\theta_p\cos(\phi_p-\phi)\Big]
\big(2\rho^2-r^2-r\Delta^{\frac{1}{2}}+B)
-L^2\cos\theta\cos\theta_p\Bigg\},\nonumber\\
\frac{{\rm d}\phi_p}{{\rm d}\lambda}&=&
\frac{\Phi}{\rho^3}\Bigg\{
     -B\frac{\sin(\phi_p-\phi)\sin\theta}{\sin\theta_p}
     +LA\Big[\frac{\cos(\phi_p-\phi)\sin 2\theta}{2\tan\theta_p}
       -\cos\theta^2\Big]+L\rho^2\Delta^{-\frac{1}{2}} \Bigg\},\\
\frac{{\rm d}\theta_p}{{\rm d}\lambda}&=&
\frac{\Phi\cos\theta}{\rho^3}\Bigg\{
     B\big[\tan\theta\cos(\phi_p-\phi)\cos\theta_p-\sin\theta_p\big]+LA\sin\theta\sin(\phi_p-\phi)\Bigg\},\nonumber
\end{eqnarray}
where 
\begin{eqnarray*}
A&=&2L\sin\theta_p\sin(\phi_p-\phi)\sin\theta+2\Delta^{\frac{1}{2}},\\
B&=&Ar+\rho^2\Delta^{-\frac{1}{2}}\left(
M-r-\Delta^{\frac{1}{2}}\right).
\end{eqnarray*}

We note the very simple form of the ordinary first order differential
equations obtained for $\dot{t}$, $\dot{r}$, $\dot{\theta}$ and
$\dot{\phi}$, where overdots denote differentiation with respect to
$\lambda$. The $\dot{\Phi}$, $\dot{\theta}_p$ and $\dot{\phi}_p$
expressions are slightly more complex but still are first order
ordinary differential equations.  This approach is therefore useful
because of the computational advantage of first order equations in
terms of efficiency and stability.  \\

An interesting aspect of the geodesic equations (\ref{eq:system1}) and
(\ref{eq:system2}) is that they do not rely on conserved quantities
such as the energy, angular momentum or Carter constant along the
photon path. As a consequence, one can actually compute these
quantities and verify that they are indeed conserved along the
trajectory. This allows one to check easily the analytical expressions
(\ref{eq:system1}) and (\ref{eq:system2}) as well as the corresponding
numerical code.  The conserved energy $E_p$ and angular momentum $J_p$
of a photon geodesic are evaluated by making use of the Killing
vectors $K_t$ and $K_{\phi}$ associated with the temporal and axial
symmetries of (\ref{eq:boyer-lindquist}) respectively. We have
\begin{equation}
\label{eq:K_t}
K_t = \frac{\Delta^{\frac{1}{2}}}{\rho}e_t-\frac{L}{\rho r}e_{\phi}.
\end{equation}
and
\begin{equation}
\label{eq:K_phi}
K_{\phi} = -\frac{L\Delta^{\frac{1}{2}}\sin^2\theta}{\rho}e_t+
            \frac{r^2+L^2}{\rho r}e_{\phi}.
\end{equation}
Therefore
\begin{equation}
E_p=p\cdot K_t=\Phi\left[\frac{L\sin\theta\sin\theta_p\sin(\phi_p-\phi)+
\Delta^{\frac{1}{2}}}{\rho}\right]
\end{equation}
and
\begin{equation}
J_p=p\cdot K_{\phi}=-\Phi\sin\theta\left[
\frac{(L^2+r^2)\sin\theta_p\sin(\phi_p-\phi)+L\Delta^{\frac{1}{2}}\sin\theta}.
{\rho}\right]
\end{equation}
The derivatives $\dot{E_p}$ and $\dot{J_p}$ with respect to the affine
parameter $\lambda$ can be obtained. As expected, substituting for the
expressions of $\dot{r}$, $\dot{\theta}$, $\dot{\phi}$, $\dot{\Phi}$,
$\dot{\theta}_p$ and $\dot{\phi}_p$ found above leads to
\begin{equation}
\frac{dE_p}{d\lambda}=\frac{dJ_p}{d\lambda}=0. 
\end{equation}

%\end{minipage}

\bsp  
\label{lastpage}
\end{document}